\documentstyle[12pt]{article}
\newcommand{\beq}{\begin{equation}}
\newcommand{\eeq}{\end{equation}}
\newcommand{\barr}{\begin{eqnarray}}
\newcommand{\earr}{\end{eqnarray}}

\newcommand{\andy}[1]{ }

\def\tt{\widetilde}

\def\cC{{\cal C}}
\def\cZ{{\cal Z}}

\begin{document}
% \draft command makes pacs numbers print
%\draft

\begin{titlepage}
\begin{flushright}
\today \\
BA-TH/97-286\\
\end{flushright}
\vspace{.5cm}
\begin{center}
{\LARGE Temporal behavior and quantum Zeno time
of an excited state of the hydrogen atom}
% repeat the \author\address pair as needed

\quad

{\large P. FACCHI and S. PASCAZIO\\
           \quad    \\

        Dipartimento di Fisica, Universit\`a di Bari

        and Istituto Nazionale di Fisica Nucleare, Sezione di Bari \\
 I-70126  Bari, Italy \\

}

\vspace*{.5cm} PACS numbers: 03.65.Bz; 31.30.Jv; 31.30.+w
\vspace*{.5cm}

{\small\bf Abstract}\\ \end{center}

{\small The quantum ``Zeno" time of the 2P-1S transition of
the hydrogen atom is computed and found to be approximately $3.59
\cdot 10^{-15}$s (the lifetime is approximately $1.595 \cdot
10^{-9}$s). The temporal behavior of this system is analyzed in a purely 
quantum field theoretical framework and is
compared to the exponential decay law.

}

\end{titlepage}

\newpage
% body of paper here
Unstable systems decay according to an exponential law. Such a law
has been experimentally verified with very high accuracy on many
quantum mechanical systems. Yet, its logical status is both subtle
and delicate, because the temporal behavior of quantum systems is
governed by unitary evolutions. The seminal work by Gamow
\cite{Gamow} on the exponential law, as well as its derivation by
Weisskopf and Wigner \cite{WW} are based on the assumption that a
pole near the real axis of the complex energy plane dominates the
temporal evolution of the quantum system. This assumption leads to
a spectrum of the Breit-Wigner type \cite{BW} and to the Fermi
Golden Rule \cite{Fermigold}. However, it is well known that a
purely exponential decay law can neither be expected for very short
\cite{MandelstTamm} nor for very long \cite{Hell} times. The domain
of validity of the exponential law is limited: the long-time power
tails and the short-time quadratic behavior are unavoidable
consequences of very general mathematical properties of the
Schr\"odinger equation \cite{temprevi}.

The short-time behavior \cite{Beskow}, in particular, turns out to be very
interesting, due to its apparently paradoxical consequences
leading the so-called quantum Zeno effect.
Recent theoretical and experimental work \cite{Cook}
has focussed on the temporal behavior of a two-level system
whose Rabi oscillations, induced by an rf field, are hindered
by another, ``measuring" field of different frequency.
It should be noticed, however, that the idea of making use of an oscillating
system to test the quantum Zeno effect is at variance with the original 
proposals, based on truly unstable systems \cite{Beskow}.
For this reason, alternative schemes were recently
proposed \cite{MPS,gaveaumodel}, that do not require any reinterpretation of
the experimental data \cite{underst}.

The purpose of this Letter is to investigate the characteristic features of
the short-time nonexponential region of a {\em truly unstable system}.
Our attention will be focussed on a transition of
the hydrogen atom: We shall endeavor to give an accurate estimate of the
``Zeno" time for this system. Our general conclusions,
however, will be valid for any two-level system interacting with a quantum
field (as far as the theory is renormalizable).

Let us start by outlining the main features of the problem.
Let $|\psi_0\rangle$ be the wave function of a given quantum system at time
$t=0$. The evolution is governed by the unitary operator
$U(t) = \exp (-iHt/\hbar)$, where $H$ is the Hamiltonian.
The ``survival" or nondecay probability at time $t$ is
the square modulus of the survival amplitude 
\andy{ndq, naiedef}
\barr
& &P(t) = |\langle \psi_0 |e^{-iHt/\hbar}|\psi_0\rangle |^2 
= 1 - t^2/\tau_{\rm Z}^2 + \cdots
\label{eq:ndq}\\
& &\tau_{\rm Z}^{-1} \equiv \frac{\triangle H}{\hbar} =
\frac{1}{\hbar}
   \left( \langle \psi_0|H^2|\psi_0\rangle -
          \langle \psi_0|H|\psi_0\rangle^2 \right)^{1/2}.
\label{eq:naiedef}
\earr
The short-time expansion is quadratic in $t$ and therefore yields a vanishing 
decay rate for $t \rightarrow 0$. This quadratic behavior is
in manifest contradiction with the exponential law
that predicts an initial nonvanishing decay rate (the inverse of 
the lifetime).
The quantity $\tau_{\rm Z}$ will be referred to as ``Zeno time,"
in the present paper.

Unfortunately,
when one considers quantum field theory, things do not work out that easily.
In the above (naive) derivation, one assumes that all moments of $H$ in the
state $|\psi_0\rangle$ are finite and (implicitly) that $|\psi_0\rangle$ is
normalizable and belongs to the domain of definition of $H$ \cite{NaPa4}.
If the volume of the box containing the system
is not finite, the spectrum of the Hamiltonian is continuous
and the Zeno time turns out to be inversely proportional to some power
of a frequency cutoff $\Lambda$:
$\tau_{\rm Z} \propto 1/\Lambda^\alpha$. This is a very
general property, essentially due to the singular nature of the product of
local observables when computed at short distances \cite{BMT}.

However, if the theory is renormalizable, this divergence can be tamed
by introducing a {\em natural} cutoff for the system.
In the present paper we shall just concentrate our attention on
such a situation: We will show that it is indeed possible to compute
the value of $\tau_{\rm Z}$ for the 2P-1S transition of the hydrogen atom.
The result is finite. This confirms that a quantum Zeno region is not a 
phenomenon peculiar to the quantum mechanics of finite systems; rather, it is 
present even in the more general framework of quantum field theory, at least 
for a renormalizable theory.

We start from the total Hamiltonian ($\hbar=c=1$)
\andy{totham2}
\barr
 H &=& H_{\rm atom} +
H_{\rm EM} + H_{\rm int}\nonumber\\ &=& \sum_{i=1}^2 E_i
|i\rangle\langle i| +
\sum_\beta
\int_0^\infty d\omega \,
\omega a^\dagger_{\omega\beta} a_{\omega\beta} \label{eq:totham2}
\\
 & &+ \sum_\beta \int_0^\infty
d\omega \left[ \varphi_\beta(\omega) a^\dagger_{\omega \beta}
|1\rangle\langle 2|
+ \varphi^*_\beta(\omega) a_{\omega \beta} |2\rangle\langle 1|
\right],\nonumber
 \earr
where the first term is the free Hamiltonian of a two-level atom,
the second term the Hamiltonian of the free EM field and the
third term the interaction Hamiltonian. We considered only the
linear part of the interaction (in the so-called rotating wave approximation)
and expanded it in the
energy-angular momentum basis for photons \cite{Heitler}, with
$\sum_\beta = \sum_{j = 1}^\infty\sum_{m=-j}^j\sum_{\lambda = 0}^1
$, where $|i\rangle
\; (i=1,2)$ are the atomic states (of energy $E_i$), $\lambda$
defines the photon parity $P=(-1)^{j+1+\lambda}$, $j$ is the total
angular momentum (orbital+spin) of the photon, $m$ its magnetic
quantum number and
 \andy{boscomm2}
 \beq
 [a_{\omega j m \lambda}, a^\dagger_{\omega' j' m'\lambda'}] =
\delta(\omega-\omega')\delta_{j j'}\delta_{m m'}\delta_{\lambda\lambda'}.
 \label{eq:boscomm2}
 \eeq
The quantities $\varphi_{\beta}(\omega)$ are the matrix elements of
the interaction Hamiltonian between the states
\andy{stati}
 \beq
 |1; 1_{\omega\beta}\rangle  \equiv |1\rangle \otimes |\omega, j, m,
\lambda\rangle, \quad
 |2; 0\rangle \equiv |2\rangle \otimes |0\rangle ,  \label{eq:stati}
 \eeq
where the first ket refers to the atom and the second to the photon.
We concentrate now our attention on
the 2P-1S transition of hydrogen:
$|1\rangle\equiv|n_1=1,l_1=0,m_1=0\rangle,\;
|2\rangle\equiv|n_2=2,l_2=1, m_2\rangle$. Conservation of angular
momentum and parity ensures the validity of the selection rules
$j=1, \;m = m_2,\; \lambda= 1$. This reduces the sum over $\beta$
in the interaction Hamiltonian to the single term $\bar{\beta} =
(1, m_2, 1)$. In this case, the matrix elements were exactly
evaluated by Moses \cite{Moses} and Seke \cite{Seke}
 \andy{phidef1}
 \barr
 \varphi_\beta (\omega) &=& \langle 1, 1_{\omega\beta}|H_{\rm int}|2,
0\rangle =
\varphi_{\bar{\beta}}(\omega)\delta_{\beta\bar{\beta}}\nonumber\\
&=& i(\chi\Lambda)^\frac{1}{2}\frac{\left(
\frac{\omega}{\Lambda}\right)^\frac{1}{2}}{\left[1 + \left(
\frac{\omega}{\Lambda}\right)^2 \right]^2}
\delta_{j1}\delta_{mm_2}\delta_{\lambda1},
 \label{eq:phidef1}
 \earr
with
 \andy{lambdachi}
 \barr
 \Lambda = \frac{3}{2} \alpha m_e \simeq 8.498\cdot 10^{18} \mbox{rad/s},
 \nonumber\\
 \chi
%=  \frac{2 \alpha}{3 \pi} |\bmv_{12}|^2
 = \frac{2}{\pi}
\left(\frac{2}{3}\right)^9 \alpha^3 \simeq 6.435 \cdot 10^{-9},
 \label{eq:lambdachi}
 \earr
where
%$\bmv_{12}=\langle 1|\bmhv |2 \rangle$, $\bmhv$ being the velocity
%operator of the electron in the atom,
$\alpha$ is the fine structure constant and $m_e$ the electron
mass. $\Lambda$ is the {\em natural} cutoff defining the
atomic form factor and taking into account all retardation effects:
in natural units, $\Lambda = 3/2a_0$, where $a_0$
is the Bohr radius, so that wavelengths shorter than $a_0$ do not contribute 
significantly to the interaction. The physical origin of $\Lambda$ is 
ascribable to the exponential behavior of the atomic orbitals,
which fall off like $\exp(-r/n a_0)$ (where $r$ is the radial coordinate and 
$n$ the principal quantum number): for the 2P-1S transition, the orbitals
overlap like $\exp(-r/a_0) \cdot \exp(-r/2 a_0) = \exp(-r \Lambda)$.
Notice that $\Lambda$ is not put ``by hand," like in analysis involving 
the dipole approximation, but naturally emerges from calculation 
\cite{Moses,Seke}.

We assume that the system is initially (at time
$t=0$) in the eigenstate $|2, 0\rangle$ of the unperturbed
Hamiltonian $H_0=H_{\rm atom} + H_{\rm EM}$, whose eigenvalue is
$\omega_0 = E_2 - E_1 = \frac{3}{8}
\alpha^2 m_e \simeq 1.550 \cdot 10^{16}$ rad/s.
%\beq
%H_0 |2, 0\rangle = \omega_0 |2, 0\rangle,
%\eeq
%where we set $E_1 = 0$ and
%\andy{omega0}
%\beq
%\omega_0 = E_2 - E_1 = \frac{3}{8}
%\alpha^2 m_e \simeq 1.550 \cdot 10^{16} \mbox{rad/s}
%\label{eq:omega0}
%\eeq
We shall set $E_1 =0$. It is now straightforward to compute the Zeno
time, according to the definition (\ref{eq:naiedef}):
\andy{tauis}
\barr
& & \frac{1}{\tau_{\rm Z}^2} = \langle 2, 0|H_{\rm int}^2|2,
0\rangle
=
\sum_\beta
\int_0^\infty d\omega \, |\langle 2, 0|H_{\rm int}|1,
1_{\omega\beta}\rangle|^2\nonumber \\
 & & =\int_0^\infty d\omega |\varphi_{\bar{\beta}}(\omega)|^2 = \chi
\Lambda^2 \int_0^\infty dx \frac{x}{(1+x^2)^4}
= \frac{\chi}{6} \Lambda^2. 
\label{eq:tauis}
 \earr
Inserting the values (\ref{eq:lambdachi}) of $\Lambda$ and $\chi$ we obtain
\andy{taudet}
\beq
 \tau_{\rm Z} = \sqrt{\frac{6}{\chi}} \frac{1}{\Lambda} = 
(3 \pi)^{\frac{1}{2}}
\left(\frac{3}{2}\right)^{\frac{7}{2}} \frac{1}{\alpha^{\frac{5}{2}} m_e}
\simeq 3.593 \cdot 10^{-15} \mbox{s}.
 \label{eq:taudet}
 \eeq
This is our first result. It is an 
estimate of the duration of the Zeno region for a truly unstable system.

Observe that for hydrogen-like atoms of nuclear charge $Z$, the 
Zeno time scales (unfavorably) like $Z^{-2}$. This is because $\Lambda\propto 
Z/a_0$ and $\chi\propto Z^2\alpha^3$.

It is also worth stressing that the value of $\tau_{\rm Z}$,
due to its very structure, would not be modified by the presence of
counter-rotating terms in the Hamiltonian (\ref{eq:totham2}), whose
contribution to (\ref{eq:tauis}) vanishes. Even the introduction of additional
atomic levels would not modify this result, within the framework of the 
rotating wave approximation [whose validity is discussed after Eq.\ (27)].
On the other hand, a straightforward but rather lengthy calculation
shows that the introduction of the
other atomic levels, whose principal quantum number is $\nu$,
{\em and} of counter-rotating terms in the interaction Hamiltonian 
$H^\prime_{\rm int}$ yields the following expression for the Zeno time:
\andy{taunew}
\beq
\frac{1}{\tau^{\prime 2}_{\rm Z}} = \langle 2, 0|H^{\prime 2}_{\rm int}|2,
0\rangle =
\sum_{\nu, \beta}
\int_0^\infty d\omega \, |\langle 2, 0|H^\prime_{\rm int}|\nu,
1_{\omega\beta}\rangle|^2 = \frac{1.4210}{\tau_{\rm Z}^2},
\label{eq:taunew}
\eeq
where the matrix elements are computed as in \cite{Seke}.
Equation (\ref{eq:taunew}) yields a 
20\% correction to the value of the Zeno time.

It is now interesting to look at the temporal behavior of our system at
longer times. There is previous related work
\cite{KnightMilonni'76,DavidNuss,Hillery,SH,Enaki} on this subject.
The survival amplitude and its Laplace transform read
 \andy{survamp}
 \barr
 y(t) &=& \langle 2, 0|e^{-i H t}|2, 0\rangle, \nonumber\\
 \tt{y}(s) &=& \int_0^\infty dt\, e^{-st} y(t) = \langle 2, 0|\frac{1}{s
+ iH}|2, 0\rangle. \label{eq:survamp}
\earr
We make use of the identity
 \barr
 \frac{1}{s+iH} &=& \frac{1}{s+iH_0} -i\frac{1}{s+iH_0}
H_{\rm int} \frac{1}{s+iH_0}\nonumber\\ & &- \frac{1}{s+iH_0}
H_{\rm int}
\frac{1}{s+iH_0} H_{\rm int}
\frac{1}{s+iH}
 \earr
and by introducing a complete orthonormal set of eigenstates of the 
unperturbed
 Hamiltonian $H_0$ [note that the interaction Hamiltonian
$H_{\rm int}$ has nonvanishing matrix elements only between the
states (\ref{eq:stati})] we easily obtain
 \andy{,qs2}
 \barr
 \tt{y}(s)=\frac{1-Q(s)\tt{y}(s)}{s+i\omega_0}\;\;\Rightarrow\;\;
 \tt{y}(s)=\frac{1}{s+i\omega_0+Q(s)},\\
 Q(s) \equiv \int_0^\infty dk\, |\varphi_{\bar{\beta}}(k)|^2 
\frac{1}{s + i k}.
 \label{eq:qs2}
 \earr
By inverting the transform we get
\andy{survamp1,qs3}
 \barr
 y(t) &=& \frac{1}{2 \pi i}
\int_{\rm B} ds \frac{e^{s \Lambda t}}{s + i
\frac{\omega_0}{\Lambda} + \chi \bar{Q}(s)},
 \label{eq:survamp1} \\
 & & \quad \bar{Q}(s) \equiv\frac{1}{\chi\Lambda}Q(s\Lambda)
 = -i \int_0^\infty dx \frac{x}{(1+x^2)^4} \frac{1}{x-is}.
 \label{eq:qs3}
 \earr
where B is the so-called Bromwich path, i.e.\ a vertical line at the
right of all the singularities of $\tt{y}(s)$, and we used (\ref{eq:phidef1}).
Notice that $Q$ and $\bar{Q}$ are self-energy contributions.
It is straightforward to integrate (\ref{eq:qs3}) to get
 \andy{qs4}
 \barr
 \bar{Q}(s) &=& \frac{-15\pi i -(88-48\pi i)s - 45\pi is^2 + 144s^3}
 {96(s^2 - 1)^4}\nonumber\\
 & &+\frac{15\pi is^4- 72s^5 - 3\pi is^6 + 16s^7-96 s\log s}{96(s^2 - 1)^4} .
 \label{eq:qs4}
 \earr
The quantity $\bar{Q}(s)$ has a logaritmic branch cut extending
from $0$ to $-i\infty$, and no singularities on the first Riemann
sheet (physical sheet). Indeed, the fourth order zeros of the
denominator $s=\pm 1$ are also zeroes of the numerator and
$\bar{Q}(\pm 1) = \frac{\pm 32 - 5\pi i}{256}$.
On the second Riemann sheet the function $\bar{Q}(s)$ becomes
 \beq
 \bar{Q}_{\rm II}(s) = \bar{Q}(s e^{-2\pi i}) = \bar{Q}(s) + 2 \pi i
\frac{s}{(s^2 -1)^4},
 \eeq
where the additional term represents the discontinuity across the
cut. It is easy to show that $\tt{y}(s)$ has a pole on the second
Riemann sheet. From the denominator in (\ref{eq:survamp1}), by
expanding $\bar{Q}_{\rm II}(s)$ around
$-i\frac{\omega_0}{\Lambda}-0^+ =
-i\frac{\alpha}{4}-0^+$, we get a power series whose convergence
radius is $\frac{\alpha}{4}$, because of the branching point at the
origin. Therefore
\andy{qspole}
 \barr
 s_{\rm pole} &=& -i\frac{\alpha}{4}-\chi\bar{Q}_{\rm II}\left(
-i\frac{\alpha}{4} -0^+\right) +O(\chi^2)\nonumber\\
&=& -i\frac{\alpha}{4}-\chi\bar{Q}\left(
-i\frac{\alpha}{4} + 0^+\right)+O(\chi^2) \\
&\equiv& -i \frac{\alpha}{4} + i \frac{\Delta E}{\Lambda} -
\frac{\gamma}{2\Lambda} ,
 \label{eq:qspole}
 \earr
because $\bar{Q}_{\rm II}(s)$ is the
analytical continuation of $\bar{Q}(s)$ below the branch cut.
%\barr
% \gamma &=& 2\chi\Lambda
%\Re\left[\bar{Q}\left(-i\frac{\alpha}{4}+0^+\right)\right]+O(\chi^2) ,
%\nonumber\\
% \Delta E &=& -\chi\Lambda
%\Im\left[\bar{Q}\left(-i\frac{\alpha}{4}+0^+\right)\right]+O(\chi^2)
% \earr
By (\ref{eq:qs3}) we get
 \andy{fgr, shift}
 \barr
 \gamma &=& 2\pi |\varphi_{\bar{\beta}}(\omega_0)|^2 + O(\chi^2)\nonumber\\
&=& 2\pi\chi\frac{\omega_0}{\left[ 1+\left(\frac{\alpha}{4}\right)^2
\right]^4}+O(\chi^2)
\simeq 6.268\cdot10^8\mbox{s}^{-1}\label{eq:fgr}\\
  \Delta E &=& P\int_0^\infty d\omega |\varphi_{\bar{\beta}}(\omega)|^2
\frac{1}{\omega -\omega_0}+O(\chi^2)\simeq 0.5 \chi \Lambda,
 \label{eq:shift}
 \earr
which are the Fermi ``Golden Rule" (yielding the lifetime
$\tau_{\rm E}=\gamma^{-1} \simeq 1.595\cdot 10^{-9}$s) and the
second order correction to the level energy $E_2$. Notice that
$\Delta E$ is not the Lamb shift, but only the shift of the $2P$
level due to its interaction with the ground state
\cite{MosesShift,Seke}. Observe that for hydrogen-like atoms of 
nuclear charge $Z$, $\tau_{\rm E} \propto (\chi \omega_0)^{-1}$ 
scales like $Z^{-4}$, so that the ratio 
$\tau_{\rm Z}/\tau_{\rm E}$ has the favorable scaling $Z^{2}$.
This might be important for experimental observation of the Zeno
region.

The exponential law is readily obtained by
deforming the original Bromwich path into a new contour
$C=C_1+C_2$, composed of a small circle $C_1$ turning anticlockwise
around the simple pole $s_{\rm pole}$ on the second Riemann sheet
and a path $C_2$ starting from $-\infty$ on the second sheet,
turning around the branch point $s=0$ and extending back to
$-\infty$ on the first sheet. We get
 \beq
 y(t) = y_{\rm pole}(t) + y_{\rm cut}(t),
 \eeq
where
\andy{poleorder}
 \barr
\;\;\;y_{\rm pole}(t) &=&
%\frac{1}{2\pi i}\int_{C_1} ds \frac{e^{s \Lambda
%t}} {s + i\frac{\omega_0}{\Lambda} + \chi \bar{Q}(s)} =
%\frac{e^{-\frac{\gamma}{2}t} e^{-i(\omega_0-\Delta E) t}}
%{1+\chi\bar{Q}'_{\rm II}(s_{\rm pole})}=
\cZ e^{-\frac{\gamma}{2} t}e^{-i(\omega_0-\Delta E)t +i\zeta} , \\
& &\quad\cZ e^{i\zeta} \equiv \frac{1}{1+\chi\bar{Q}'_{\rm
II}(s_{\rm pole})}
= 1 + O(\chi)
\label{eq:poleorder}
 \earr
and the prime denotes derivative.
Notice that $\chi=O(\alpha^3)$ and $\zeta \simeq -2.02\cdot10^{-8}$.
As is well known, the exponential law is obtained by neglecting the 
contribution arising from the branch cut. Let us estimate the 
latter. From (\ref{eq:survamp1})
\andy{cutcontr}
\barr
y_{\rm cut}(t) &=& \frac{1}{2\pi i} \int_{C_2} ds\, e^{s\Lambda
t}\left[
\frac{1}{s+i\frac{\omega_0}{\Lambda}+\chi\bar{Q}(s)}
\right]\nonumber\\
& = &  \chi x_0^2 \int_0^\infty d\xi \frac{ \xi e^{-\xi} 
(\xi^2 x_0^2-1)^{-4}}{ \left[\xi x_0-i\frac{\omega_0}{\Lambda}-
\chi\bar{Q}(-\xi x_0)\right]
\left[\xi x_0-i\frac{\omega_0}{\Lambda}-\chi\bar{Q}_{\rm II}
(-\xi x_0)\right]}, \nonumber \\
\label{eq:cutcontr}
 \earr
where $x_0 = 1/t\Lambda$. 
At times $t \gg \Lambda^{-1}$ (so that $x_0 \rightarrow 0$)
\andy{cutcontry}
 \barr
 y_{\rm cut}(t) \sim \chi x_0^2 \frac{\int_0^\infty d\xi
\, \xi e^{-\xi}}{
\left[-i\frac{\omega_0}{\Lambda}-\chi\bar{Q}(0)
\right]^2} = -\chi \frac{\cC}{(\omega_0 t)^2}, \\
\cC \equiv \left(1-\frac{5}{8}\pi\frac{\chi}{\alpha}\right)^{-2}
=1 + O(\chi).
 \label{eq:cutcontry}
 \earr
The expressions (\ref{eq:qs4})-(\ref{eq:cutcontry}) 
are robust against the introduction of counter-rotating terms in the 
Hamiltonian (\ref{eq:totham2}), which 
can be shown to contribute only first-order corrections in 
$\chi =O(\alpha^3)$ in (\ref{eq:qs4}) and therefore second-order 
corrections
in (\ref{eq:qspole}), (\ref{eq:poleorder}) and (\ref{eq:cutcontry}).

Summarizing, the general expressions (valid $\forall t \geq 0$)
for the survival amplitude $y(t)$
and survival probability $P(t)=|y(t)|^2$ are respectively
\andy{geny,genP}
\barr
y(t) & = & \cZ e^{-\frac{\gamma}{2} t} e^{-i(\omega_0-\Delta E)t +i\zeta}
-\chi \frac{\cC}{(\omega_0 t)^2}h(t)e^{i\eta(t)}
\label{eq:geny} \\
P(t) & = & \cZ^2 e^{-\gamma t}+ \chi^2\frac{\cC^2}{(\omega_0
t)^4}h^2(t)\nonumber\\
 & & - 2 \chi\frac{\cC\cZ}{(\omega_0 t)^2} e^{-\frac{\gamma}{2} t} h(t)
 \cos\left[(\omega_0-\Delta E) t +\eta(t)-\zeta \right] ,
\label{eq:genP}
\earr
where $h(t)$ and $\eta(t)$ are real functions satisfying
\barr
& &\lim_{t\rightarrow 0}\frac{h(t)}{(\omega_0 t)^2} =
\frac{\sqrt{1+\cZ^2-2\cZ\cos\zeta}}{\chi\cC}\qquad
\lim_{t\rightarrow\infty}h(t)=1,\nonumber\\
& &
\eta(0)=\arctan\left(\frac{\cZ\sin\zeta}{\cZ\cos\zeta-1}\right)\qquad
\lim_{t\rightarrow\infty}\eta(t)=0.
\earr
Notice the presence of an oscillatory term, in Eq.\ (\ref{eq:genP}).
Physically, it represents an interesting (fully quantum mechanical) 
interference effect between the cut and the pole contribution to the survival
amplitude (\ref{eq:geny}).

For short and long times, Eq.\ (\ref{eq:genP}) yields
 \andy{shortt,largetimes1}
 \barr
P(t) \! &\sim& \! 1 - \frac{t^2}{\tau_{\rm Z}^2} \qquad (t\ll \tau_{\rm Z})
 \label{eq:shortt} \\
P(t) \! &\sim& \! \cZ^2 e^{-\gamma t}+ \chi^2\frac{\cC^2}{(\omega_0 t)^4} 
- 2 \chi\frac{\cC\cZ}{(\omega_0 t)^2} e^{-\frac{\gamma}{2} t}
 \cos\left[(\omega_0-\Delta E) t -\zeta \right] . \label{eq:largetimes1} \\
  & & \qquad\qquad\qquad\qquad\qquad\qquad\qquad\qquad(t\gg \Lambda^{-1})
  \nonumber
 \earr
Numerical investigation of (\ref{eq:genP}) shows that 
the ``long-time" expansion is already valid for rather short times
$t \;^>_\sim \; 2\cdot 10^{-17}\mbox{s}$. For even shorter times, 
the system undergoes a 
rapid initial oscillation, of duration about $200 \cdot \Lambda^{-1}
\simeq  2.3\cdot 10^{-17}\mbox{s}$, and then quickly relaxes towards the 
asymptotic expression (\ref{eq:largetimes1}). The initial convexity of the
curve is given by (\ref{eq:shortt}), which agrees extremely well with the 
numerical investigation.

The above analysis clarifies an important point:
in contrast with a widespread, naive expectation, the short time
behavior, yielding a vanishing decay rate, is nothing but the first
of a series of oscillations, whose amplitude vanishes exponentially
with time, eventually leading to a power law. The asymptotic frequency
of the oscillations is essentially $\omega_0$ [see (\ref{eq:largetimes1})]:
any correction (like our $\Delta E$, or the total 
Lamb shift, or fine structure effects, not considered in our analysis) is 
at most of order $10^{-6}\omega_0$. The transition to a power law occurs when
the two summands in (\ref{eq:geny}) are comparable, so that
$(\omega_0 t)^2e^{-\frac{\gamma}{2}t}\approx\chi$, namely for
$t\simeq 98$ lifetimes \cite{DavidNuss,Hillery,SH,KnightMilonni'76}.

The above conclusions, derived for the hydrogen atom in interaction
with the EM field, are generally valid for a renormalizable (or
superrenormalizable) theory: Any interaction Hamiltonian of the
type (\ref{eq:totham2}), which does {\em not} contain derivative
couplings in the fields, yields similar results. One should notice,
however, that the evaluation of the duration of the Zeno region
depends on the frequency cutoff: in general, one expects a
dependence on some inverse power of $\Lambda$ \cite{BMT}; for example, in our
case $\tau_{\rm Z} = O(\Lambda^{-1})$, as in (\ref{eq:taudet}). An
accurate estimate of $\Lambda$ can pose in general a difficult
problem.

An interesting problem is to understand whether the initial quadratic behavior
(\ref{eq:shortt}) is experimentally observable. This is an experimentally 
challenging task, that raises interesting 
theoretical and experimental questions 
about the problem of state preparation. 
The time scales involved are very small, so that a sharp 
initial state preparation, even by modern pulsed-laser techniques,
appears difficult. On the other hand, state preparation by means of 
{\em indirect} excitation processes, e.g.\ by electron or 
ion collision, seems more realistic. 

It is also worth stressing that
the problem of sharply defining the initial moment of excitation might 
be circumvented: close scrutiny of 
Eqs.\ (\ref{eq:geny})-(\ref{eq:largetimes1})
suggests that experimental observation of the probability 
oscillations would not only provide a direct evidence of the cut 
contribution to the survival amplitude, but also an indirect, 
yet convincing, proof of the presence of the Zeno region, in the light of the 
discussion following (\ref{eq:largetimes1}).
%
%\vspace*{.5cm}
%{\bf Acknowledgements} \vspace{.3cm}
%
%We would like to thank ***** for helpful discussions.
%%%%%%%%%%%%%%%%%%%%%%%%%%%%%%%%%%%%%%%%%%%%%%%%%%%%%%%%%%%%%%%%%
% now the references. delete or change fake bibitem. delete next three
%   lines and directly read in your .bbl file if you use bibtex.

% figures follow here
% \begin{figure}
% \caption{}
% \label{}
% \end{figure}
%%%%%%%%%%%figure%%%%%%%%%%%%%%%%%%%%
%\begin{figure}[t]
%\vspace*{6cm}
%\epsfig{file=zeno.eps, width=5.5cm}
%\caption{Behavior of the survival probability $P(t)$.
%For illustrative purposes, we set 
%$\chi=0.1, \gamma=0.1$s$^{-1}$ and $\omega_0=3$s$^{-1}$. Different scales have
%been used for 
%the two graphs: The frequency of the oscillations is practically the same
%over the entire time domain [see Eq.\ (\ref{eq:largetimes1})].
%The dashed line is the exponential and the dotted line the power law.
%}
%\label{fig:behavior}
%\end{figure}
%%%%%%%%%%%%%%%%%%%%%%%%%%%%%%%%%%%%%%
% tables follow here
%
% Here is an example of the general form of a table:
% Fill in the caption in the braces of the \caption{} command. Put the label
% that you will use with \ref{} command in the braces of the \label{} command.
% Insert the column specifiers (l, r, c, d, etc.) in the empty braces of the
% \begin{tabular}{} command.
%
% \begin{table}
% \caption{}
% \label{}
% \begin{tabular}{}
% \end{tabular}
% \end{table}


\begin{thebibliography}{000}
\bibitem{Gamow} \andy{Gamow}
G. Gamow, Z. Phys. {\bf 51}, 204 (1928).

\bibitem{WW} \andy{WW}
V. Weisskopf and E.P. Wigner, Z. Phys. {\bf 63}, 54 (1930);
{\bf 65}, 18 (1930).

\bibitem{BW} \andy{BW}
G. Breit and E.P. Wigner,
Phys. Rev. {\bf 49}, 519 (1936).

\bibitem{Fermigold} \andy{Fermigold}
E. Fermi, {\it Nuclear Physics} (Univ.
Chicago, Chicago, 1950) pp.\ 136, 148; 
%See also {\it Notes on Quantum
%Mechanics. A Course Given at the University of Chicago in 1954} edited by E
%Segr\'e (Univ.\ Chicago, Chicago, 1960) Lec.\ 23; 
Rev. Mod. Phys. {\bf 4}, 87 (1932).

\bibitem{MandelstTamm} \andy{MandelstTamm}
L. Mandelstam and I. Tamm, J. Phys. {\bf 9}, 249 (1945);
V. Fock and N. Krylov, J. Phys. {\bf 11} 112 (1947).

\bibitem{Hell} \andy{Hell}
E.J. Hellund, Phys. Rev. {\bf 89}, 919 (1953);
M. Namiki and N. Mugibayashi, Prog. Theor. Phys. {\bf 10}, 474 (1953).
L.A. Khalfin, Dokl. Acad. Nauk USSR {\bf 115}, 277 (1957)
[Sov. Phys. Dokl. {\bf 2}, 340 (1957)];
Zh. Eksp. Teor. Fiz. {\bf 33}, 1371 (1958)
[Sov. Phys. JETP {\bf 6}, 1053 (1958)].

\bibitem{temprevi}\andy{temprevi}
H. Nakazato, M. Namiki and S. Pascazio,
Int. J. Mod. Phys. {\bf B10}, 247 (1996).

\bibitem{Beskow} \andy{Beskow}
A. Beskow and J. Nilsson, Arkiv f\"ur Fysik
{\bf 34}, 561 (1967);
L.A. Khalfin, Zh. Eksp. Teor.
Fiz. Pis. Red. {\bf 8}, 106 (1968) [JETP Letters {\bf 8}, 65 (1968)];
L. Fonda, G.C. Ghirardi, A. Rimini and T. Weber,
Nuovo Cim. {\bf A15}, 689 (1973); {\bf A18}, 805 (1973);
%A. DeGasperis, L. Fonda and G.C. Ghirardi, Nuovo Cim. {\bf A21}, 471 (1974);
B. Misra and E.C.G. Sudarshan, J. Math. Phys. {\bf 18}, 756 (1977);
A. Peres, Am. J. Phys. {\bf 48}, 931 (1980); Ann. Phys. {\bf 129}, 33 (1980).

\bibitem{Cook} \andy{Cook}
R.J. Cook, Phys. Scr. {\bf T21}, 49 (1988); W.H. Itano, D.J. Heinzen, J.J.
Bollinger and D.J. Wineland, Phys. Rev. {\bf A41},
2295 (1990).
%Phys. Rev. {\bf A43}, 5168 (1991); 
%T. Petrosky, S. Tasaki and I. Prigogine, Phys. Lett. {\bf A151}, 109 (1990);
%Physica {\bf A170}, 306 (1991); A. Peres and A. Ron, Phys. Rev.
%{\bf A42}, 5720 (1990); L.E. Ballentine, Phys. Rev. {\bf A43}, 5165
%(1991); V. Frerichs and A. Schenzle, in {\it Foundations of Quantum
%Mechanics} edited by T D Black {\em et al} (World Scientific,
%Singapore, 1992); S. Inagaki, M. Namiki and T. Tajiri, Phys. Lett.
%{\bf A166}, 5 (1992); D. Home and M.A.B. Whitaker, J. Phys. {\bf
%A25}, 657 (1992); Phys. Lett. {\bf A173}, 327 (1993); Ph. Blanchard
%and A. Jadczyk, Phys. Lett. {\bf A183}, 272 (1993); S. Pascazio, M.
%Namiki, G. Badurek and H. Rauch, Phys. Lett. {\bf A179}, 155
%(1993).

\bibitem{MPS} \andy{MPS}
E. Mihokova, S. Pascazio and L.S. Schulman,
Phys. Rev. {\bf A56}, 25 (1997).

\bibitem{gaveaumodel} \andy{gaveaumodel}
L.S. Schulman, J. Phys. {\bf A30}, L293 (1997);
B. Gaveau and L.S. Schulman, J. Stat. Phys. {\bf 58}, 1209 (1990);
J. Phys. {\bf A28}, 7359 (1995);
L.S. Schulman, A. Ranfagni and D. Mugnai, Phys. Scr. {\bf 49}, 536 (1994).

\bibitem{underst} \andy{underst}
S. Pascazio and M. Namiki,
Phys. Rev. {\bf A50}, 4582 (1994);
H. Nakazato, M. Namiki, S. Pascazio and H. Rauch,
Phys. Lett. {\bf A217}, 203 (1996).

\bibitem{NaPa4} \andy{NaPa4}
H. Nakazato and S. Pascazio,
Mod. Phys. Lett. {\bf A10}, 3103 (1995).

\bibitem{BMT} \andy{BMT}
C. Bernardini, L. Maiani and M. Testa,
Phys. Rev. Lett. {\bf 71}, 2687 (1993).

\bibitem{Heitler} \andy{Heitler}
W. Heitler, Proc. Cambridge Phil. Soc. {\bf 32}, 112 (1936);
%V.B. Berestetskii, Zh. Eksp. Teor. Fiz. {\bf **}, *** (19**)  \fin\
%[Sov. Phys. JETP {\bf 17} (1947) 12];
A.I. Akhiezer and V.B. Berestetskii,
{\it Quantum electrodynamics} (Interscience Publ., New York, 1965).

\bibitem{Moses} \andy{Moses}
H.E. Moses,
Lett. Nuovo Cimento {\bf 4}, 51 (1972);
Phys. Rev. {\bf A8}, 1710 (1973).

\bibitem{Seke} \andy{Seke}
J. Seke, Physica {\bf A203}, 269; 284 (1994).

\bibitem{KnightMilonni'76} \andy{KnightMilonni'76}
P.L. Knight and P.W. Milonni, Phys. Lett. {\bf 56A}, 275 (1976).

\bibitem{DavidNuss} \andy{DavidNuss}
L. Davidovich and H.M. Nussenzveig, in
{\it Foundations of radiation theory and quantum electrodynamics},
ed. A.O. Barut (Plenum, New York, 1980), p.83.

\bibitem{Hillery} \andy{Hillery}
M. Hillery, Phys. Rev. {\bf A24}, 933 (1981).

\bibitem{SH} \andy{SH}
J. Seke and W. Herfort, Phys. Rev. {\bf A40}, 1926 (1989);
J. Seke, Phys. Rev. {\bf A45}, 542 (1992).

\bibitem{Enaki} \andy{Enaki}
N.A. Enaki, Zh. Eksp. Teor. Fiz. {\bf 109}, 1130 (1996)
[Sov. Phys. JETP {\bf 82}, 607 (1996)].

\bibitem{MosesShift} \andy{MosesShift}
H.S. Hoffman and H.E. Moses,
Lett. Nuovo Cimento {\bf 4}, 54 (1972).

%\bibitem{protdec} \andy{protdec}
%L. A. Khalfin Phys. Lett. {\bf 112B}, 223 (1982);
%C.B. Chiu, B. Misra and E.C.G.  Sudarshan,
%Phys. Lett. {\bf 117B}, 34 (1982);
%L. Fonda, G.C. Ghirardi and T. Weber,
%Phys. Lett. {\bf 131B}, 309 (1983);
%L.P. Horwitz and E. Katznelson,
%Phys. Rev. Lett. {\bf 50}, 1184 (1983);
\end{thebibliography}
\end{document}